\documentclass[11pt]{article}
\usepackage[margin=1in]{geometry}
\usepackage{graphicx}

\def\chinoonepmninotwo{\ensuremath{\mathchoice%
      {\displaystyle\raise.4ex\hbox{$\displaystyle\tilde\chi^\pm_1\tilde\chi^0_2$}}%
         {\textstyle\raise.4ex\hbox{$\textstyle\tilde\chi^\pm_1\tilde\chi^0_2$}}%
       {\scriptstyle\raise.3ex\hbox{$\scriptstyle\tilde\chi^\pm_1\tilde\chi^0_2$}}%
 {\scriptscriptstyle\raise.3ex\hbox{$\scriptscriptstyle\tilde\chi^\pm_1\tilde\chi^0_2$}}}~}

\def\ninotwo{\ensuremath{\mathchoice%
    {\displaystyle\raise.4ex\hbox{$\displaystyle\tilde\chi^0_{2}$}}%
    {\textstyle\raise.4ex\hbox{$\textstyle\tilde\chi^0_{2}$}}%
    {\scriptstyle\raise.3ex\hbox{$\scriptstyle\tilde\chi^0_{2}$}}%
    {\scriptscriptstyle\raise.3ex\hbox{$\scriptscriptstyle\tilde\chi^0_{2}$}}}}

\def\ninoone{\ensuremath{\mathchoice%
    {\displaystyle\raise.4ex\hbox{$\displaystyle\tilde\chi^0_{1}$}}%
    {\textstyle\raise.4ex\hbox{$\textstyle\tilde\chi^0_{1}$}}%
    {\scriptstyle\raise.3ex\hbox{$\scriptstyle\tilde\chi^0_{1}$}}%
    {\scriptscriptstyle\raise.3ex\hbox{$\scriptscriptstyle\tilde\chi^0_{1}$}}}}

\def\chinoonepm{\ensuremath{\mathchoice%
      {\displaystyle\raise.4ex\hbox{$\displaystyle\tilde\chi^\pm_1$}}%
         {\textstyle\raise.4ex\hbox{$\textstyle\tilde\chi^\pm_1$}}%
       {\scriptstyle\raise.3ex\hbox{$\scriptstyle\tilde\chi^\pm_1$}}%
 {\scriptscriptstyle\raise.3ex\hbox{$\scriptscriptstyle\tilde\chi^\pm_1$}}}}
\def\chinoonemp{\ensuremath{\mathchoice%
      {\displaystyle\raise.4ex\hbox{$\displaystyle\tilde\chi^\mp_1$}}%
         {\textstyle\raise.4ex\hbox{$\textstyle\tilde\chi^\mp_1$}}%
       {\scriptstyle\raise.3ex\hbox{$\scriptstyle\tilde\chi^\mp_1$}}%
 {\scriptscriptstyle\raise.3ex\hbox{$\scriptscriptstyle\tilde\chi^\mp_1$}}}}

\newcommand{\ifb}{\mbox{fb\(^{-1}\)}}

\newcommand{\Ht}{\ensuremath{ m_\mathrm{eff}^{3\ell} }}
\newcommand{\Hboost}{\ensuremath{ H^{\mathrm{boost}} }}

\newcommand{\pTsoft}{\ensuremath{  p_\mathrm{T}^{\mathrm{soft}} }}

\newcommand{\RMetJets}{\ensuremath{ R\left(E_\mathrm{T}^\mathrm{miss},\mathrm{jets}\right) }}

\newcommand{\pt}{\ensuremath{  p_\mathrm{T} }}
\newcommand{\met}{\ensuremath{  E_\mathrm{T}^\mathrm{miss} }}

\newcommand{\pTmiss}{\ensuremath{\mathbf{p}_\mathrm{T}^\mathrm{miss}}}

\newcommand{\HTppthreeone}{\ensuremath{H^{\mathrm{PP}}_{\mathrm{T}\ 3,1}}}
\newcommand{\pTpp}{\ensuremath{p^\mathrm{PP}_\mathrm{T}}}
\newcommand{\pTcm}{\ensuremath{p^\mathrm{CM}_\mathrm{T}}}

\newcommand{\pTI}{\ensuremath{p^\mathrm{I}_\mathrm{T}}}

\def\Title#1{\begin{center} {\Large {\bf #1} } \end{center}}
\def\Author#1{\begin{center} {\normalsize {\sc #1} } \end{center}}
\def\Institution#1{\begin{center} {\normalsize {\it #1} } \end{center}}
\def\Abstract#1{\noindent {\normalsize {\bf Abstract:} {\normalfont #1}}}
\def\Conference{\vspace{4mm}\begin{raggedright} {\normalsize {\it Talk presented at the 2019 Meeting of the Division of Particles and Fields of the American Physical Society (DPF2019), July 29--August 2, 2019, Northeastern University, Boston, C1907293.} } \end{raggedright}\vspace{4mm}}

\begin{document}

%
%

\Title{Search for chargino-neutralino production using an emulated recursive jigsaw reconstruction technique in three-lepton final states with the ATLAS detector}

\Author{Elodie Resseguie}

\Institution{Lawrence Berkeley National Laboratory\\ One Cyclotron Road, Berkeley, CA 94720, USA}

\Abstract{A search for supersymmetry through the pair production of electroweakinos is presented in a three-lepton final state. The analyzed proton-proton collision data taken at a centre-of-mass energy of $\sqrt{s} = 13$ TeV were collected between 2015 and 2018 by the ATLAS experiment at the Large Hadron Collider, corresponding to an integrated luminosity of $139~\ifb$. The search emulates the recursive jigsaw reconstruction technique using conventional analysis variables, searching for low-mass chargino-neutralino pair production that decays to on-shell $W$ and $Z$ bosons. The technique is validated and the excess seen previously in 2015 and 2016 data is studied while incorporating new data.}

\Conference

%
%

\section{Introduction}

\begin{figure}[h!]
  \begin{center}
    \includegraphics[width=0.25\textwidth]{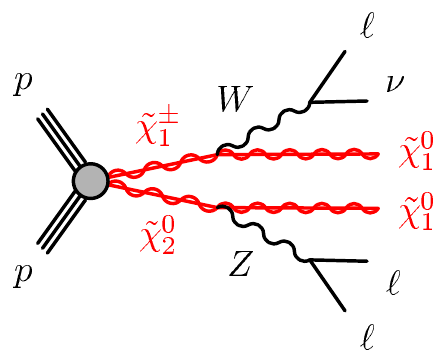}
    \includegraphics[width=0.26\textwidth]{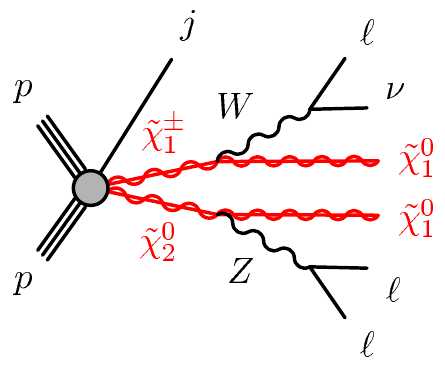}
  \end{center}
  \caption{Diagrams for the production of $\tilde\chi_1^{\pm}\tilde\chi_2^{0}$ decaying via $W$ and $Z$ bosons to three leptons and missing transverse energy in $pp$ collisions. The diagram on the right is the production $\tilde\chi_1^{\pm}\tilde\chi_2^{0}$ in association with an initial-state-radiation jet, labelled ``j".}
   \label{fig:rjr_diagrams}
\end{figure}

The search for the chargino-neutralino (\chinoonepmninotwo) pair-production with mass splitting near the electroweak scale is presented.
The targeted decay chain is shown in Figure~\ref{fig:rjr_diagrams}, with the chargino and neutralino decaying to the invisible LSP \ninoone\ and either a $W$ or $Z$ gauge boson, respectively. The \chinoonepm\ and \ninotwo\ are assumed to be purely wino and mass degenerate, and decay with 100$\%$ branching ratio to $W$ and $Z$ bosons. The \ninoone\ LSP is assumed to be pure bino. 
Both the $W$ and $Z$ bosons decay leptonically via SM branching ratios, leading to a final state with three leptons and missing momentum from two \ninoone\ and a neutrino.
The presence of initial state radiation (ISR) may lead to jets in the final state and boost the \chinoonepmninotwo\ system, enhancing the signature of the missing momentum.
The search targets a range of \chinoonepm/\ninotwo\ masses 
between 100 GeV and 450 GeV and mass splittings with respect to the \ninoone\ LSP, $\Delta m = m({\chinoonepm/\ninotwo}) - m({\ninoone})$, larger than the $Z$ boson mass.

Two ATLAS searches targeted that phase space using data collected in 2015 and 2016, corresponding to $36.1~\ifb$ of data collected, one with laboratory frame variables~\cite{Aaboud:2018jiw}, which does not see an excess of observed events above the background prediction, and another using the Recursive Jigsaw Reconstruction (RJR) technique~\cite{1607.08307,1705.10733}, which found excesses of three-lepton events in two overlapping regions, one targeting low-mass resonances and another utilizing ISR to target resonances with mass differences with respect to the LSP close to the $Z$ boson mass.

This new, independent analysis explores the intersection between the conventional and RJR approaches to better understand the tension in the exclusion limits produced by the two analyses.
It emulates the variables used by the RJR technique with conventional laboratory frame discriminating variables, providing a simple set of variables that are easily reproducible. This technique reproduces the RJR excesses in the low-mass region and ISR regions in the laboratory frame
using the same $36.1~\ifb$ of $pp$ collision data collected between 2015 and 2016 by the ATLAS detector at the LHC. 
The object and region definitions using these new emulated Recursive Jigsaw Reconstruction (eRJR) variables are kept as close as possible to those from Ref.~\cite{Aaboud:2018sua}. The excess observed in both RJR and eRJR in the 2015-2016 dataset is followed up using the eRJR technique using a larger dataset corresponding to $139~\ifb$ of $pp$ collision data collected between 2015 and 2018~\cite{ATLAS-CONF-2019-020}. 

\section{Event Selection}
The signal regions are split into two different topologies: SR-low, the low-mass region that requires a jet veto, and SR-ISR, the ISR region that requires at least one jet.

SR-low requires the $\pt$ of the first, second, and third leptons to be greater than 60 GeV, 40 GeV, and 30 GeV, respectively. Tight selection thresholds on the eRJR variables further reduce the $WZ$ contribution in the signal region.

SR-ISR requires an ISR jet, which boosts the final state objects in the same direction, enhancing the $\met$ in the final state. As a result, this SR has a requirement of $\met \geq 80$ GeV. The $\pt$ requirement on the three leptons is relaxed to be greater than 25 GeV, 25 GeV, and 20 GeV, such that the dilepton triggers are fully efficient. Additional requirements on eRJR variables further select the boosted topology of the event and that the majority of transverse momentum along the jet axis is carried by the invisible particles and not by the high-$\pt$ leptons from the $WZ$ background.

The eRJR technique emulates the RJR variables using minimal assumptions on the mass of the invisible system and calculates all kinematic variables in the laboratory frame. Some of the eRJR variables, with original RJR variable names from Ref.~\cite{Aaboud:2018sua} in parenthesis, are defined as,
\begin{itemize}
\item \met (\pTI), the $\pt$ of the invisible particles is emulated as the magnitude of the missing transverse momentum.
\item \pTsoft (\pTcm), the transverse momentum in the center-of-mass (CM) frame where the ISR system recoils against the system containing the leptons and the missing energy, is emulated as the $\pt$ of the vector sum of the four-momenta of the signal jets, leptons, and \pTmiss,
\item \pTsoft (\pTpp), the transverse momentum in the center-of mass frame of the protons (PP), is emulated as the $\pt$ of the vector sum of the four-momenta of the signal leptons and \pTmiss, being identical to \pTsoft  except for the jet veto applied.
\item \Ht (\HTppthreeone), the scalar sum of the $\pt$ of the signal leptons and the invisible system (neutrino and LSPs) in the rest frame of the sparticle pair, is emulated as the scalar sum of the $\pt$ of the signal leptons and $E_\mathrm{T}^\mathrm{miss}$.
\end{itemize}

Example correlations for the signal point of $m(\tilde{\chi}^\pm_1/\tilde{\chi}^0_2),m(\tilde{\chi}^0_1)=(200,100)$ GeV are shown for a loose selection that does not include any requirements on eRJR varibles in Figure~\ref{fig:rj_trans_corr_presel_isr_sig}. A strong correlation is observed indicating that these are very similar variables

\begin{figure}[h]
  \centering
 \includegraphics[width=0.48\textwidth]{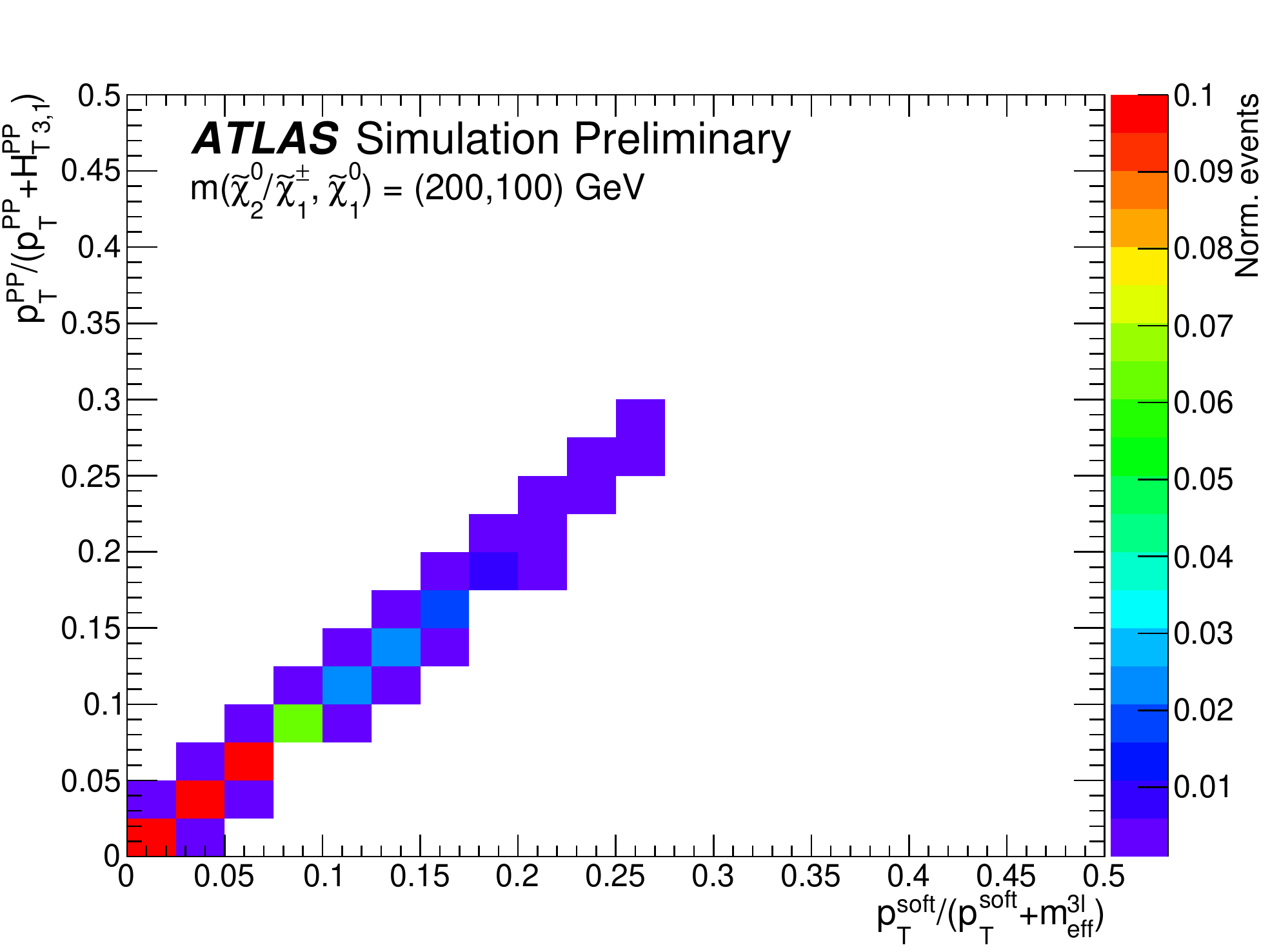}
  \caption{Correlations of an RJR (y-axis) and eRJR (x-axis) variable for the $m(\tilde{\chi}^\pm_1/\tilde{\chi}^0_2),m(\tilde{\chi}^0_1)=(200,100)$ signal point.}
  \label{fig:rj_trans_corr_presel_isr_sig}
\end{figure}

\section{Background Validation}

The dominant irreducible background is $WZ$ production which is estimated from Monte Carlo simulation whose yields are normalized to data using control regions (CR).
Other irreducible backgrounds include $ZZ$, $VVV$, $ttV$, and Higgs processes, and are estimated directly from MC simulation due to their small contribution.
Reducible backgrounds, containing at least one fake lepton, are estimated using data-driven methods for $Z$+jets and $Z$+$\gamma$ processes, and estimated from MC normalized in a CR for top-quark like processes $t\bar{t}$, $Wt$, and $WW$.

Four validation regions, VR-low, VR-ISR, VR-ISR-small $\pTsoft$, VR-ISR-small $\RMetJets$, are designed to check the agreement of the background estimation with data in regions kinematically closer to the signal regions (SR), typically targeting the extrapolation from CR to SR of a specific variable. The VR definitions are also chosen to keep signal contamination below 10$\%$.
Figure~\ref{fig:VR} shows example distributions in VR-low and VR-ISR-small $\RMetJets$ for the full background prediction. 
There is generally good agreement seen between the expected background prediction and the observed data.

\begin{figure}[h]
  \centering
      \includegraphics[width=0.45\textwidth]{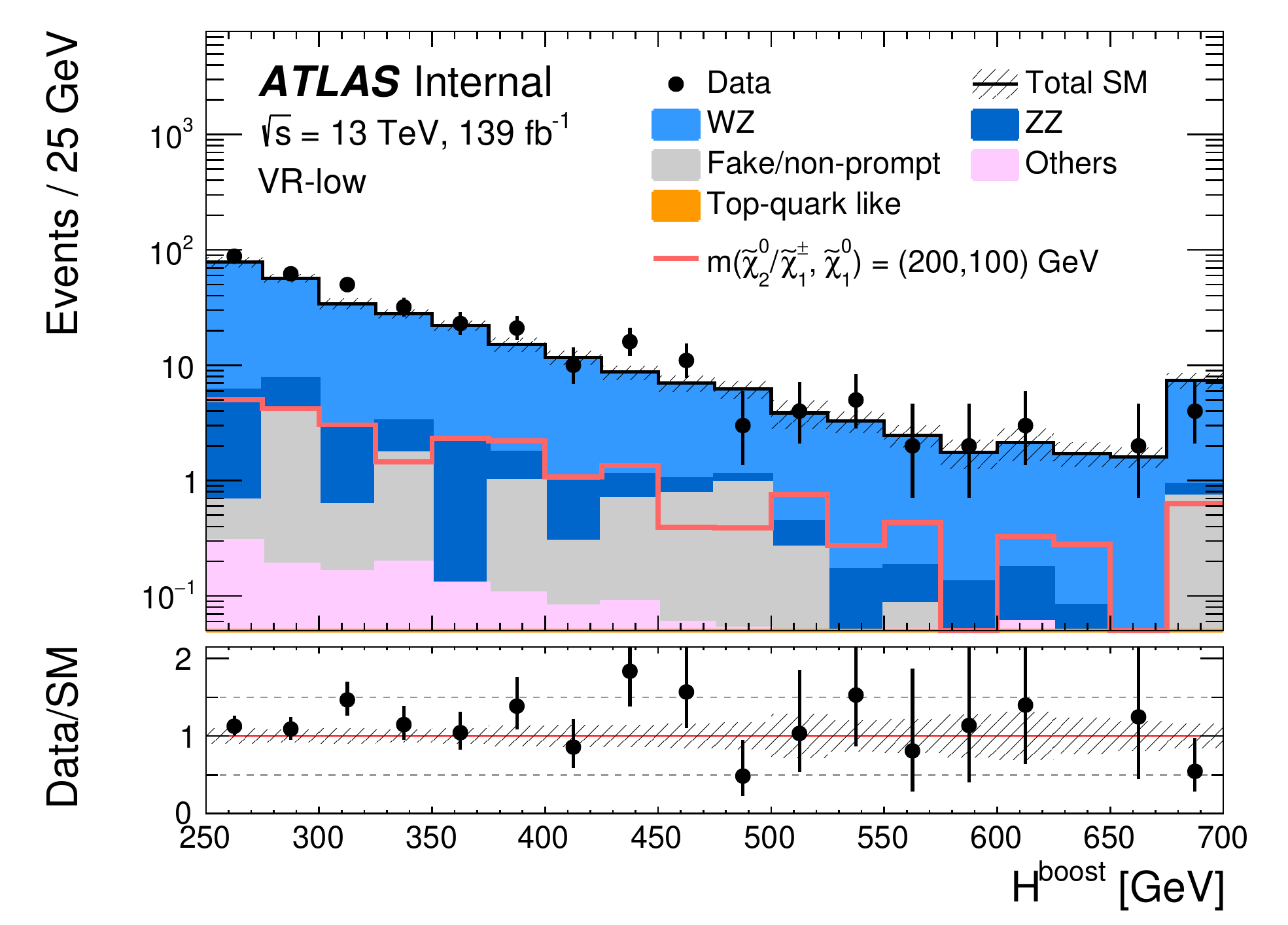}
      \includegraphics[width=0.45\textwidth]{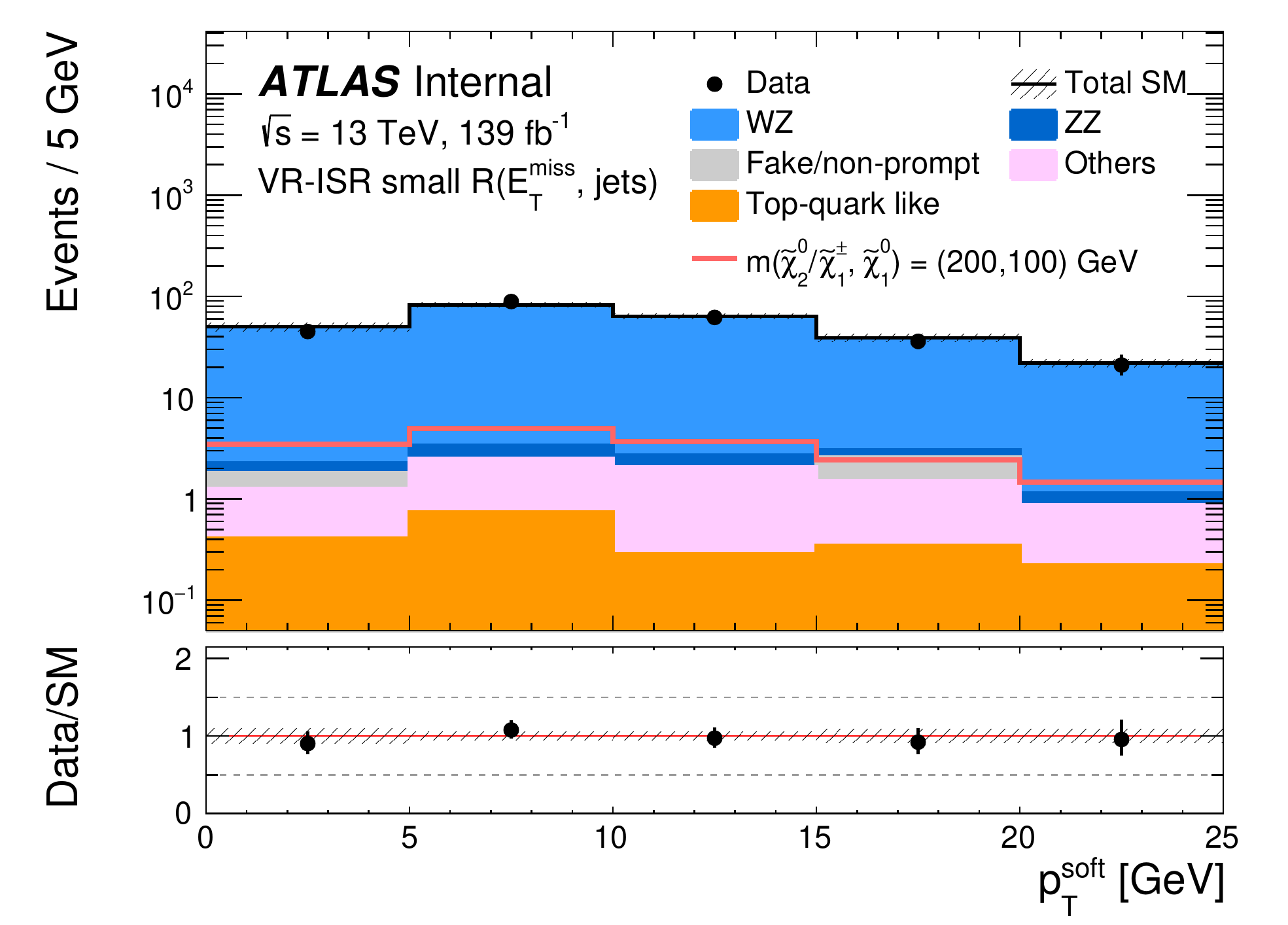}

    \caption{Kinematic distributions showing the data and post-fit background in VR-low for $\Hboost$ (left) and VR-ISR-small-$\RMetJets$ for $\pTsoft$ (right).}
  \label{fig:VR}
 \end{figure}

\section{Results}

No significant excess of data above the background prediction is observed. As a result, model-independent limits are derived at 95$\%$ confidence level for each SR and summarized in Table~\ref{table.results.exclxsec.pval.upperlimit.SR}. Limits on $\sigma_\mathrm{vis}$ are set at 0.16~fb in SR-low and 0.13~fb in SR-ISR.

\begin{table}
\centering
\begin{tabular*}{\textwidth}{@{\extracolsep{\fill}}lcccccc}
\noalign{\smallskip}\hline\noalign{\smallskip}
{Signal channel}    & $N_\mathrm{obs}$   & $ N_\mathrm{exp}$   & $\sigma_\mathrm{vis}$[fb]  &  $S_\mathrm{obs}^{95}$  & $S_\mathrm{exp}^{95}$  & $p(s=0)$ ($Z$)  \
\\
\noalign{\smallskip}\hline\noalign{\smallskip}
 SR-low    & 51 & 46 $\pm$ 5 &    $0.16$ &  $22.0$ & $ { 20.7 }^{ +6.2 }_{ -4.3 }$& $ 0.27$~$(0.60)$   \\%
\noalign{\smallskip}\hline\noalign{\smallskip}
 SR-ISR    & 30 & $23.0 \pm 2.2$  & $0.13$ &  $17.8$ & $ { 12.1 }^{ +5.3 }_{ -2.0 }$ & $ 0.10$~$(1.27)$\\%
\noalign{\smallskip}\hline\noalign{\smallskip}
\end{tabular*}
\caption[Breakdown of upper limits.]{
Summary of the model-independent limits for SR-low and SR-ISR with expected and observed yields,  the visible number of observed ($S^{95}_\mathrm{obs}$) and expected ($S^{95}_\mathrm{exp}$) events, and the discovery $p$-value and Gaussian significance $Z$ assuming no signal.
\label{table.results.exclxsec.pval.upperlimit.SR}}
\end{table}

The expected and observed exclusion contours as a function of the signal $\chinoonepm/\ninotwo$ and LSP $\ninoone$ masses are shown in Figure~\ref{fig:pull_plot_summary_yields_isr}.
Masses can be excluded when the $Z/W$ bosons of the decay are on mass-shell, such that the mass splittings $\Delta m$ are close to or larger than the $Z$ boson mass.
Signal $\chinoonepm/\ninotwo$ are excluded for masses up to 350 GeV for small $\ninoone$ masses in which $\Delta m$ is large.

\begin{figure}[h!]
    \centering
        \includegraphics[width=0.55\textwidth]{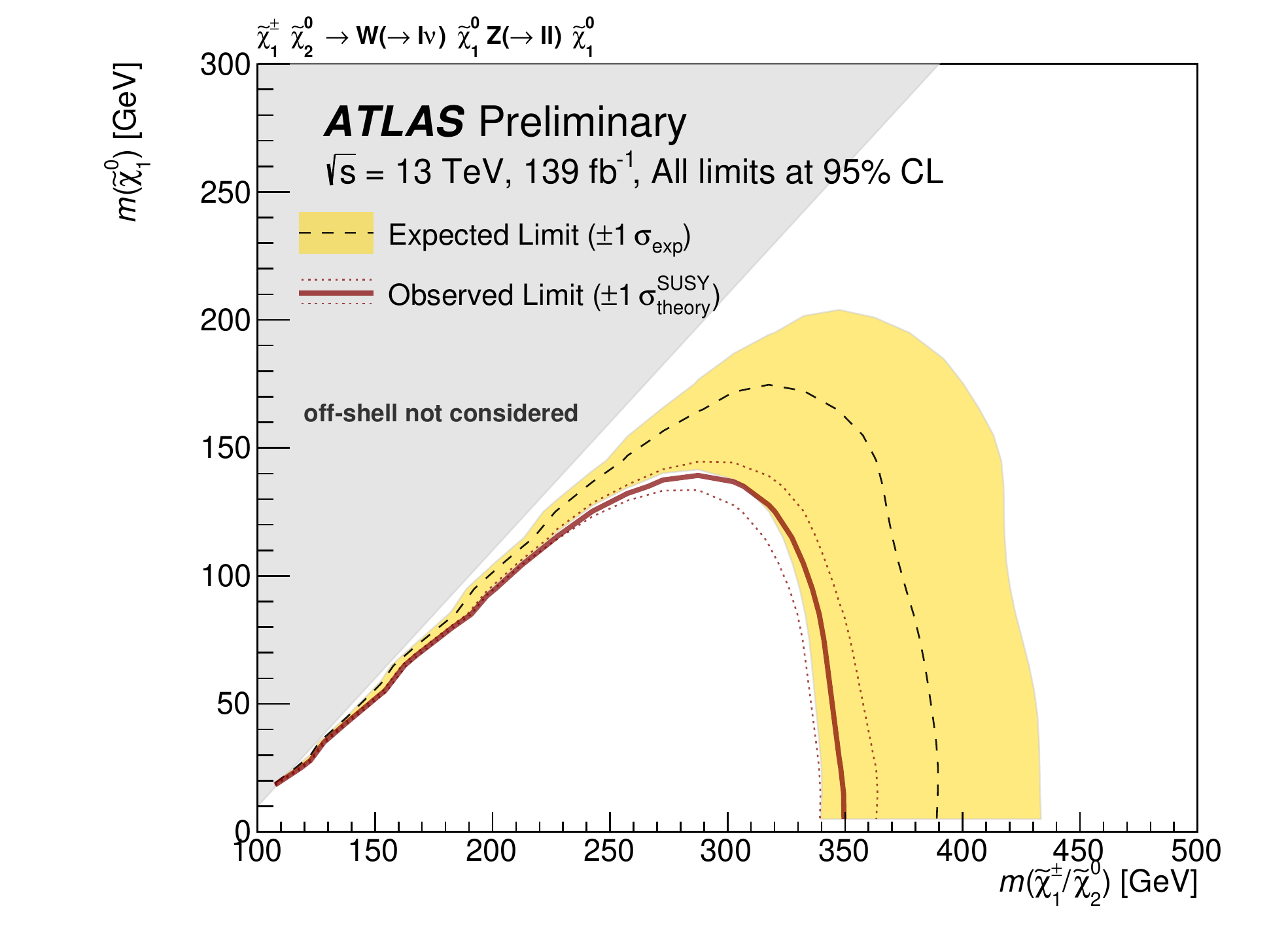}
    \caption{
    Expected (dashed blue) and observed (solid red) exclusion contours on \chinoonepmninotwo production assuming on-shell $W/Z$ decays
    as a function of the \chinoonepm/\ninotwo\ and \ninoone\ masses, and derived from the combined fit of low-mass and ISR regions.
    The yellow band reflects the $\pm 1 \sigma$ uncertainty on the expected limits due to uncertainties in the background prediction and experimental uncertainties affecting the signal.
    The dotted red lines correspond to the $\pm 1 \sigma$ cross section uncertainty of the observed limit derived by varying the signal cross section within its uncertainty.
    }
    \label{fig:pull_plot_summary_yields_isr}
\end{figure}


\clearpage

\begin{thebibliography}{99}


\bibitem{Aaboud:2018jiw}
ATLAS Collaboration, Search for electroweak production of supersymmetric particles in final
states with two or three leptons at $\sqrt{s} = 13$ TeV with the ATLAS detector, Eur. Phys. J. {\bf C}78
(2018) no. 12, 995

\bibitem{1607.08307}
P. Jackson, C. Rogan, and M. Santoni, Sparticles in motion : Analyzing compressed SUSY
scenarios with a new method of event reconstruction, Phys. Rev. D {\bf 95} (2017) 035031

\bibitem{1705.10733}
P. Jackson and C. Rogan, Recursive Jigsaw Reconstruction: HEP event analysis in the
presence of kinematic and combinatoric ambiguities, Phys. Rev. D {\bf 96} (2017) 112007

\bibitem{Aaboud:2018sua}
ATLAS Collaboration, Search for chargino-neutralino production using recursive jigsaw
reconstruction in final states with two or three charged leptons in proton-proton collisions at 
$\sqrt{s} = 13$ TeV with the ATLAS detector, Phys. Rev. D {\bf 98} (2018) no. 9, 092012

\bibitem{ATLAS-CONF-2019-020}
ATLAS Collaboration, Search for chargino-neutralino production with mass splittings near
the electroweak scale in three-lepton final states in $\sqrt{s} = 13$ TeV $pp$ collisions with the
ATLAS detector, Tech. Rep. ATLAS-CONF-2019-020, CERN, Geneva, May, 2019.


\end{thebibliography}
\end{document}